\documentclass[11pt]{article}
\usepackage{amsmath, amssymb,amsxtra}
\usepackage{slashed}
\usepackage{verbatim}
\usepackage{latexsym}
\usepackage{euscript}
\usepackage{amsfonts}
\usepackage{verbatim}
\usepackage{cancel}
\usepackage{soul}
\usepackage[normalem]{ulem}
\usepackage[]{array}
\usepackage[]{mathrsfs}
\usepackage[utf8]{inputenc}
\usepackage[T1]{fontenc}
\usepackage{makeidx}
\makeindex
\usepackage{graphicx}
\usepackage{amsmath}
\usepackage{amssymb}
\usepackage{amsxtra}
\usepackage{color}
\def\be{\begin{eqnarray}}
\def\ee{\end{eqnarray}}
\def\0{\nonumber}
\def\d{\partial}

\def\sfg{{\sf g}}

\usepackage{slashed}
\usepackage{verbatim}
\usepackage{latexsym}\usepackage{color}
\usepackage{euscript}
\usepackage[curve]{xypic}

\newcommand\ED{\EuScript{D}}

\def\sfg{{\sf g}}

\normalfont\large
\begin{document}
\vskip 1cm
\begin{flushright}
{SISSA/26/2026/FISI}\\
\end{flushright}
\vskip 1cm

\begin{center}
 {\LARGE Massless neutrinos in a left-right symmetric universe}

\vskip 1cm
 {\large  L.~Bonora\footnote{email:bonora@sissa.it},\\
 International School for Advanced Studies (SISSA),\\Via
Bonomea 265, 34136 Trieste, Italy}

\end{center}
\vskip 2cm
{\bf Abstract}
This comment is a comment stimulated by the 1957 Goldhaber et al. experiment, that detected a massless left-handed electron neutrino with negative helicity. The standard description of this neutrino is by means of a Weyl spinor. It is shown below that this description is incomplete, and in fact one needs a massless Dirac spinor to comply with the requirements of quantum field theory. A Dirac spinor is composed of two Weyl spinors. This in turn raises the necessity to explain the fate of the right-handed electron neutrino, partner of the Goldhaber et al. experiment's one. A possible explanation for its invisibility is suggested here in terms of a left-right symmetric extension of the standard model.

\section{Introduction}\label{s:intro}

In this comment we would like to return to 1958 and discuss the outcome of the Goldhaber et al. experiment, \cite{Goldhaber}. The authors, Maurice Goldhaber, Lee Grodzins and Andrew W. Sunyar, used the electron capture in Europium atoms and their subsequent decay into unstable Samarium plus an electron neutrino to measure the helicity of the latter. They concluded that the emitted neutrino has negative helicity. This means in particular that the neutrino is a massless particle with definite chirality (left) and negative helicity. We believe that it represents a serious challenge to the standard 
description of massless neutrinos, interpreted as Weyl spinors. In recent articles, \cite{bonorasoldati2019,LBRS26}, we have called attention on the fact that Weyl spinors are mathematical objects to which we can associate energy, momentum and definite helicity, but they in general lack a propagator. In other words they are well defined classical objects but ill defined quantum objects. If this is correct, what are the mathematical entities that represent the neutrino of the Golhaber et al. experiment? In this comment we would like to point out a possible explanation for this apparent puzzle.

We suggest to view the experiment against the backdrop of a recently proposed enlarged standard model. This model consists of two sector, the left sector is the minimal standard model (MSM), the right sector is a mirror of the MSM with reversed chiralities for all the fermions, \cite{BG24}. The two sectors share the same weak and gravitational interactions, but have distinct strong and electromagnetic forces. In \cite{LB25,LB26} it has been proposed to interpret the right sector as describing the dark matter world. In this model there is room for interpreting the Goldhaber et al. neutrino as a  massless Dirac fermion, made of a left-handed Weyl spinor plus a right-handed one. The catch is to explain why we do not see the right-handed component. And this is what the enlarged SM cou;d explain.

The Goldhaber et al. experiment after the discovery of neutrino oscillations, \cite{Fukuda}, raised another well known question: neutrino oscillations are usually explained with the existence of massive neutrinos, which may be either Dirac or Majorana (the literature is very large, there are various reviews on this subject, here we limit ourselves to quote few of them: \cite{Wyler83,mohapatra06,Bilenky07,King07,Dvornikov11,Kayser18,Vitiello21}). But this may be in contradiction with the existence of the massless neutrino of Goldhaber experiment. Neutrino oscillations require the presence in the model of more than one family and perhaps even a larger number of neutrinos than those standard of the MSM.  In this paper we do not aim to analyze and explain this issue. We limit ourselves to recall that the present experimental evidence on neutrino oscillations is not be in contradiction with the existence of massless neutrinos (in this case the electron neutrino) and that various models of neutrino oscillations actually contain massless neutrinos, \cite{Bilenky07}. Based on this
we assume that, whatever is the extension of our model we refer to, it admits a massless neutrino, which we identify with the electron neutrino of the Goldhaber experiment.
 
 The paper is organized as follows. Section 2 contains a summary description of the Goldhaber et. experiment. Section 3 is devoted to a lightning discussion about massless fermions. In section 4 the left-right extension of the MSM alluded to above is presented in a considerably shortened version. Finally section  5 is devoted to our main argument, the possible explanation of the puzzle concerning the nature of the  massless electron neutrino of the Goldhaber et al. experiment.

\section{The Goldhaber et al. experiment}

In a very clever experiment, deservedly famous, in 1957 Goldhaber, Grodzins and Sunyar, \cite{Goldhaber}, were able to measure the helicity of the electron neutrino and showed that it is negative. The experiment consists in the capture of an orbital electron with the emission of a neutrino, i.e. the reaction is
\be
p^+ +e^-\longrightarrow n+\nu_e\label{pennu}
\ee
More precisely the initial configuration is an atom of Europium, $_{63}Eu^{152}$ in the state $\bf 0^-$; the electron capture gives rise to an excited isotope of Samarium, $_{62} Sm^{*152} $,  plus an electron neutrino. The  $_{62} Sm^{*152} $ decays into the stable $_{62} Sm^{152} $ plus a $\gamma$. Now, owing to the spin 0 state of the Europium, the neutrino plus the Samarium spins must equal the absorbed electron spin, which is $\frac 12$. Therefore the spin of $_{62} Sm^{*152} $ and that of neutrino must be opposite in order to equal the latter. But they move in opposite directions in order to conserve the momentum. Thus the neutrino and the $_{62} Sm^{*152}$ have the same polarization.

It follows that in order to know the neutrino polarization one must measure the $_{62} Sm^{*152}$'s one. This measurement is possible because the $_{62} Sm^{*152}$ decays into the  $_{62} Sm^{152}$ (which is spin 0) and a photon. Thus their two spins must point in the same direction, due to momentum conservation since they are both $\bf 1^-$ states. If the photon is emitted in the same direction as the  $_{62} Sm^{*152}$ it will have the same polarization, thus the same polarization as the electron neutrino. 

The experiment of considerable complexity is based on the previous chain of events. Based on it the author were able to verify that the emitted photons were  polarized and with negative helicity. Consequently the they concluded that the also  electron neutrino helicity is negative.

This experiment has an enormous importance, which has perhaps not been adequately appreciated or simply ignored. To start with it means that the electron neutrino coming out of this experiment is massless. The discovery of the neutrino oscillations, \cite{Fukuda}, has raised a puzzle. Although the neutrino oscillations by themselves are not enough to exclude that there are massless neutrinos, it is not simple to explain that electron neutrinos are massless in this experiment while they participate for instance in the phenomenon of solar neutrino oscillations. But several models have been constructed that reconcile neutrino oscillations with the existence of massless neutrinos. But this is a more serious puzzle created by the Goldhaber experiment, because of the definite helicity/chirality of the electron neutrino. For it raises a theoretical question: how do we represent massless neutrinos with definite  chirality? The standard  answer to this question is: by means of Weyl spinors. The trouble is that classically (free) Weyl spinors obey a Weyl-Dirac equation, carry definite momentum, energy and chirality, have null mass, but do not have a causal propagator: a fatal obstacle for quantization. The usual attitude in the literature is to replace this nonexistent propagator with the Dirac one. But the latter calls into question also the opposite chirality. In other words, in order to represent the left-handed negative-chiral electron neutrino of the Goldhaber experiment we must call on a right-handed patner. Which has never been seen. 

It should be mentioned that a not unpopular idea exists to overcome the above difficulties and consists in making recourse to Majorana neutrinos. Majorana neutrinos can carry a mass, therefore they can explain oscillations and have a well-defined propagator, the same as the Dirac one. But precisely for this reason they consist of two functionally related partners with opposite chirality, with the consequence that, even if they are massless, the helicity is not well defined, see \cite{LBRS26}. Therefore if they may very well explain oscillations, they are of no help in explaining the Goldhaber experiment.

At present an acceptable explanation of the Goldhaber experiment to our best knowledge is still lacking. In this paper we would like to see if an explanation may be contained in the extension of the SM put forward in \cite{BG24,LB25,LB26}. But before coming to that proposal we have to clarify the terms of the problem, in particular the concepts of massless fermions, Weyl fermions and their difference with Dirac and Majorana fermions.

\section{Massless fermions in 4d}

An extensive introduction to Dirac, Weyl and Majorana fermions in 4 dimensions is contained in \cite{bonorasoldati2019} and a more complete one in \cite{LBRS26}. Here we present a summary of the points relevant for our discussion.

A (free) massless Dirac spinor is a four component complex spinor satisfying the Dirac equation
\be
i\slashed {\partial} \psi(x)=0, \quad\quad \slashed {\partial}= \gamma^\mu \partial_\mu\label{masslesDirac}
\ee
The Lagrangian is 
\be
{\cal L}_D= i \overline \psi\slashed{\partial} \psi, \quad\quad \overline \psi= \psi^\dagger \gamma_0\label{lagDirac}
\ee
The propagator for a Dirac fermion is the inverse of the kinetic operator $i\slashed{\partial} $, which exists and in momentum representation gives rise to the propagator $\frac i{\slashed{ p}+ i \epsilon}$.

A left/right handed Weyl spinor is obtained from $\psi$ by projecting it with the chiral projectors ${\mathbb P}_-/{\mathbb P}_+$, where ${\mathbb P}_\mp = \frac {1\mp \gamma_5}2$:
$\psi_\mp(x)= {\mathbb P}_\mp \psi(x)$.

The Lagrangian for a left handed spinor is
\be
{\cal L}_-= i \overline \psi_-\slashed{\partial} \psi_-\label{lagWeyl}
\ee
The kinetic operator (also  referred to as Dirac-Weyl operator) maps right handed component of $\psi$ to left handed ones. Therefore it can be represented as
\be
{\slashed \partial}_- = \slashed \partial\, {\mathbb P}_-\label{DWop}
\ee
It is evident that, due to the projector factor, it is impossible to invert this operator. Consequently one is lead to conclude that the (perturbative) propagator of a Weyl fermion does not exist. This is of course a dramatic problem for (perturbative) quantization, which is aggravated by the fact that one cannot define the partition function of a (free) Weyl fermion. This is due to the fact that the partition function of a free fields can be represented by the product of the eigenvalues of the kinetic operator. But for the Dirac-Weyl operator case the eigenvalue problem is not even defined since the domain and the codomain are disconnected vector spaces (apart from 0).

One might bring into question this drastic conclusion by remarking the a Weyl spinor is a two (complex) component spinor and an inverse of the kinetic operator might exist in the two component formalism. It is important to clarify this point. To this end it is convenient to use the Weyl representation for the gamma matrices.
\be
\gamma^0 =
\left\lgroup\begin{array}{cc}
0 &  {\bf 1}\\
{\bf 1} & 0\\
\end{array}\right\rgroup\qquad\quad
\gamma^{\,k} =
\left\lgroup\begin{array}{cc}
0 & \sigma_k\\
-\,\sigma_k  & 0\\
\end{array}\right\rgroup\qquad\quad
\gamma_{5} = \left\lgroup\begin{array}{cc}
- \mathbf{1} &  0\\
0 & \mathbf{1}\\
\end{array}\right\rgroup\0
\ee
where $\sigma_k$ are the Pauli matrices. In such a case therefore
\be
\psi_{-}(x)=\mathbb{P}_{-}\,\psi(x)=
\left\lgroup
\begin{array}{c}
\psi_L(x) \\ 0
\end{array}
\right\rgroup, \quad\quad
\psi_{+}(x)=\mathbb{P}_{+}\,\psi(x)=
\left\lgroup
\begin{array}{c}
0 \\ \psi_{R}(x)
\end{array}
\right\rgroup\label{psimp}
\ee
and the Dirac equation splits into two disconnected equations, with kinetic operators
\begin{equation}
\sigma\cdot\partial=\sigma^{\mu}\partial_{\mu}=\partial_{0}-\sigma_{k}\partial_{k} \equiv \partial_{L}
\qquad\quad {\rm and} \quad\quad
\tilde\sigma\cdot\partial=\tilde\sigma^{\mu}\partial_{\mu}=\partial_{0}+\sigma_{k}\partial_{k} \equiv \partial_{R},\label{Weyldiffop} 
\end{equation}
respectively. They apply to the two-component Weyl spinors $\psi_L$ and $\psi_R$, respectively. Thus we have two disconnected two-component Dirac-Weyl equations
\be
\sigma\cdot\partial \psi_L=0, \quad\quad \tilde\sigma\cdot\partial\psi_R=0 \label{DWeq}
\ee

It is possible to split also the Hamiltonian $H_0= \gamma^0 \gamma^k  p^k$, into two distinct Hamiltonians $H_R$ and $H_L$
\be
H_R= \vec{ p} \cdot \vec { \sigma} = -H_L\label{HLHR}
\ee
and find solutions (spin states) of the previous equations that satisfy
\begin{eqnarray}
&& H_L\,u_L({\bf p})=\hslash c|{\bf p}|\,u_L({\bf p})\qquad\quad
{\sf h} u_L({\bf p})= -\,\textstyle\frac12\,u_L({\bf p})
\\
&& H_L\,v_L({\bf p})=-\,\hslash c|{\bf p}|\,v_L({\bf p})\qquad\quad
{\sf h} v_L({\bf p})=\textstyle\frac12\,v_L({\bf p})\label{HLuv}
\ee
and 
\be
&& H_R \,u_R({\bf p})=\hslash c|{\bf p}|\,u_R({\bf p})\qquad\quad
{\sf h} u_R({\bf p}) = \textstyle\frac12\,u_R({\bf p})
\\
&& H_R\,v_R({\bf p})= -\,\hslash c|{\bf p}|\,v_R({\bf p})\qquad\quad
{\sf h} v_R({\bf p})=-\,\textstyle\frac12\,v_R({\bf p})\label{HRuv}
\end{eqnarray}
where 
\be
{\sf h}= \frac 12 c\, \hslash \frac {\vec {p}}{|\vec {p}|}\cdot \vec {\sigma}\label{helicity}
\ee
represents the helicity operator. A generic $\psi_L$ or $\psi_R$ can be expanded on the basis of these solutions., for instance
\be
\psi_R(x^0,{\bf x})= \int \frac {d{\bf p}} {(2\pi)^3|{\bf p}|}\left[u_R ({\bf p})e^{-i|{\bf p}| x^0 +i {\bf p}\cdot {\bf x}}+v_R ({\bf p})e^{i|{\bf p}| x^0 -i {\bf p}\cdot {\bf x}}\right]\label{psiL}
\ee

In \cite{LBRS26} it is shown that
\be
S_c(x-y)\equiv \langle 0|T \psi_R(x) \psi_R^\dagger(y)|0\rangle= i\, \partial_x \cdot \sigma \, D_0^c(x-y) \label{Sc}
\ee
where
\be
D_0^c(x-y)= \frac i{(2\pi)^4} \int d^4p \, \frac {e^{ip(x-y)}}{p^2+i\varepsilon}\label{D0c}
\ee
This leads to 
\be
\tilde \sigma \cdot \partial S_c(x-y)= \delta^{(4)}(x-y) \label{sigmadSc}
\ee
which means that the right-handed differential operator $\overline \sigma \cdot \partial$ is inverted by the two by two matrix causal Green function  $S_c(x-y)$.

Similarly
\be
\overline S_c(x-y)\equiv \langle 0|T \psi_L(x) \psi_L^\dagger(y)|0\rangle= i\, \partial_x \cdot \tilde\sigma \, D_0^c(x-y) \label{Sc}
\ee
and
\be
 \sigma \cdot \partial \overline S_c(x-y)= \delta^{(4)}(x-y) \label{sigmadbarSc}
\ee
For clarity let us rewrite the so obtained equations in the more suggestive form:
\begin{equation}
S_{c}(x-y)  =  i\partial_{L}\,D^{\,c}_0(x-y)\qquad\quad \partial_{R} S_{c}(x-y)= \delta^{\, (4)}(x-y)
\end{equation}
\begin{equation}
\bar S_{c}(x-y) = i\partial_{R}\,D^{\,c}_0(x-y)
\qquad\quad
\partial_L\,\bar S_{c}(x-y) =\,\delta^{\, (4)}(x-y)
\end{equation}

Now consider, for instance, the classical action integral for the generic left-handed Weyl spinor, $\psi_L$
\begin{equation}
S_L=\int \mathrm{d}^4x\,\psi_{L}^{\dagger}(x)\,i\sigma\cdot \partial\,\psi_{L}(x)
=\int \mathrm{d}^4x\,\psi_{L}^{\dagger}(x)\,i\partial_{L}\psi_{L}(x)\label{SL}
\end{equation}
The inverse of the causal Green's function $ \langle 0\vert T\,\psi_L(x)\,\psi_L^{\dagger}(y)\vert0\rangle $ for a left-handed Weyl spinor (i.e. the would be Weyl propagator)
is not the inverse of the kinetic operator $ i\partial_{L} $ appearing in the corresponding classical action integral but of the kinetic operator
$ i\partial_{R} $ appearing in the classical action integral of a right-handed Weyl spinor and \textit{viceversa}. This means that the process of inverting the (two-component) kinetic operator
for a left handed Weyl fermion ends up involving also the right-handed companion. It is impossible to obtain a propagator for a Weyl fermion without intruding upon the other chirality. In addition it can be proved, \cite{LBRS26}, that the notion itself of chirality is not preserved in the two-component formalism. In other words the two-component formalism for Weyl spinors is not trustworthy.

In order to obtain a trustworthy causal propagator of a massless spinor there is no other possibility but turn to Dirac (four-component) spinor.
Indeed consider the $4\times 4$ matrix
\begin{footnotesize}
\begin{eqnarray*}
S^{\,c}_D(x-y) &=& \langle0\vert T \psi(x)\,\bar{\psi}(y)\vert0\rangle
=\left\lgroup
\begin{array}{cc}
0 & \bar{S}_c(x-y)
\\
S_c(x-y) & 0
\end{array}
\right\rgroup
\\
&=&\left\lgroup
\begin{array}{cc}
0 & i\partial\cdot\bar{\sigma}
\\
i\partial\cdot \sigma & 0
\end{array}
\right\rgroup
D_{0}^{\,c}(x-y)\label{Diracprop}
\end{eqnarray*}
\end{footnotesize}
and apply $i\slashed \partial$ 
\begin{eqnarray}
i\partial\!\!\!/S^{\,c}_D(x-y) =
(-1)\left\lgroup
\begin{array}{cc}
\square & 0
\\
0 & \square
\end{array}
\right\rgroup
D_{0}^{\,c}(x-y)
=i\delta^{\, (4)}(x-y)
\end{eqnarray}
where $\square$ is the D'Alembert operator.
As expected, this proves that that the massless Schwinger propagator $ S^{(c)}_D(x-y) $ is the inverse of the massless Dirac differential operator
$ \partial\!\!\!/ $.  It is fully causal forward and backward in time. Its Fourier representation is well known and is given by
\begin{equation}
S^{\,c}_0 (x-y) = \frac{i}{(2\pi)^4}\int \mathrm{d}^4p\;\frac{ p\!\!\!/ }{p^{\,2} + i\varepsilon}\;
e^{-\,ip\cdot ( x - y )}\label{DiracpropFT}
\end{equation}

Returning to the initial hope on the two-component formalism we conclude that the latter does not improve the situation concerning the existence of a Weyl fermion propagator: a (free) Weyl propagator represented by a causal two-point Green function does not exist.

So far we have been discussing about the existence of a propagator for a free Weyl fermion, whose Lagrangian is given by eq.\eqref{lagWeyl}. What about the same problem if Weyl fermions interact (minimally) with gauge fields or a metric? This has also been extensively treated in \cite{LBRS26}, and previously in \cite{LB,BG24}. This is typically the case of the SM, possibly coupled to gravity, see below for explicit formulas. In all these theories the existence of the fermion propagators is linked to a cohomological condition dictated by the family's index theorem of Atiyah and Singer, \cite{atiyahsinger1984}. This theorem gives the index (that is, the difference between the kernel and cokernel) of the (Euclidean) Dirac-Weyl operator in terms of characteristic classes, which are specific polynomials of the gauge curvature $F$, represented by the field strenght $F_{\mu\nu}$, and the metric curvature $R$, represented by the Riemann tensor $R_{\mu\nu\lambda\rho}$. As long as the index is non-vanishing the Dirac-Weyl operator is not invertible, thus the propagator cannot exist.  The remarkable aspect of it is that the characteristic classes that  feature in the index are in one-to-one correspondence with the consistent gauge and gravitational anomalies of the theory. From the field theory point of view it is well known that such anomalies are a hazard for the theory, in that they put in jeopardy its renormalization and unitarity. The theorem establishes a correspondence between such anomalies and the non-invertibility of the Dirac-Weyl operator: as long as this type of anomalies is present in the theory the Dirac-Weyl operator is non-invertible, as a consequence the corresponding propagator does not exist. It is understandable that they represent an insurmountable obstacle to quantization. For this reason such anomalies have also been called obstructive.

What matters for our present discussion is that the vanishing of the index given by the family's index theorem is a necessary condition for the existence of the Weyl propagator. This is the case for instance for the MSM. The minimal SM is formulated in terms of a multiplet of fermions, see below, coupled to the $SU(3)\times SU(2) \times U(1)$ gauge fields. The choice of fields and group representations are such that the corresponding index of the kinetic operator vanishes, and consequently also the anomalies are absent. Then the necessary condition for the existence of the Dirac-Weyl propagators is satisfied. The question is now: how do these propagators concretely take form? The answer is rather simple: the field of the MSM rearrange themselves in such a way that only Dirac kinetic terms appear in the kinetic part of the action of the MSM. Of course, in this way all the propagators exist in the form of Dirac propagators. It must be said that we do not know if this is the general case, that is we do not know if the only way a Weyl fermion can have a propagator in a non-trivial background\footnote{When the background is trivial  the family's index is trivially vanishing, and we have seen above that in such a case the Weyl propagator does not exist.} is to be accompanied by a Weyl fermion of opposite chirality so as to form a Dirac fermion. But we consider it very likely.

After this overview of Weyl fermions the question of the previous section remains: what is the electron neutrino of the Goldhaber experiment? All the quantum numbers of that neutrino points to a Weyl fermion. But its propagator as an isolated particle does not exist. It seems that a causal propagator can exist only if it is accompanied by another particle of opposite chirality, a right-handed neutrino, which has not been experimentally detected so far. This puzzle can have perhaps an explanation within an enlarged version of the SM, which we are going to present and motivate in the next section.

\section{A left-right symmetric universe}

In \cite{BG24} and \cite{LB25} a left-right symmetric enlargement of the MSM has been proposed, motivated by its coupling with gravity. We start this section with a concise description of  this model of matter minimally coupled to gravity. The fermion matter sector, which we shall cal conventionally left, is based on the same multiplet as the MSM. In the usual SM notation it is\footnote{Following the standard notation, henceforth symbols such as $\psi_L$ and $\psi_R$  refer to spinors in the four component formalism.)}
\be
\begin{matrix} {\sf G}/fields & \quad SU(3)\quad &\quad SU(2)\quad &\quad U(1)\quad\\
\left( \begin{matrix} u \\ d\end{matrix} \right)_{\! L} & 3&2&\frac 16\\
{u_R} &  3 &1&\frac 23\\
{d_R} & 3 &1&=\frac 13\\
\left( \begin{matrix} \nu_e \\ e\end{matrix} \right)_{\! L} & 1&2&-\frac 12\\
{e_R} &  1 &1&-1
\end{matrix}\label{Lspectrum}
\ee
where $ X^c$ represents the conjugate spinor of $X$, i.e. $X^c=\gamma_0 C X^\ast$, where $C$ is the charge conjugation matrix. This multiplet couples  to the $SU(3)_L\times SU(2)\times U(1)_L$ gauge fields and to a gravity metric and connection. The explicit kinetic part of the action is
\be
S_f^{(-)}\equiv S_{fL}
&=&\int d^4 x \, \left(\sqrt{g}\, i\overline {{\psi}}  
\gamma^a
 e_a^{\mu}
\left(\ED^{(-)}_\mu+\frac 12 \omega_\mu\right)\psi\right)( x)\label{fermionaction-}
\ee
where $\psi$ represents the left multiplet \eqref{Lspectrum}, and
\be
\ED^{(-)}_\mu=\partial_\mu -i{\sfg}_X^- X^{(-)}_\mu -i{\sfg}_WW_\mu -i{\sfg}_B^-B_\mu^{(-)} \label{XWQ-}
\ee
According to the standard notation $ \omega_\mu=\omega_\mu^{ab}\Sigma_{ab}$ represents the spin connection corresponding to the metric $g$ and $\Sigma_{ab}$ the anti-hermitean Lorentz generators. 
The symbols $X^{(-)}_\mu,W_\mu,B^{(-)}_\mu$ refer to the $SU(3)_{L}, SU(2)$ and $U(1)_{L}$ potentials, respectively, and take values in the corresponding Lie algebras, whose generators are  Hermitean.  Each potential has its own distinct coupling to the fermions, which are shown explicitly in eq.\eqref{XWQ-}.
In the multiplet \eqref{Lspectrum} the doublet $\left(\begin{matrix} u\\ d\end{matrix}\right)_L$  describes left-handed particles and right-handed antiparticles, while the singlets $u_R$ and $d_R$ represent right-handed particles and left-handed antiparticles, and similarly for the leptons.

In our theory there is also a sector, which we shall call right, which represents of course the main difference with the MSM. It is characterized by opposite chiralities with respect to the MSM. The right multiplet is
\be
\begin{matrix} {\sf G}/fields & \quad SU(3)\quad &\quad SU(2)\quad &\quad U(1)\quad\\
\left( \begin{matrix} u' \\ d'\end{matrix} \right)_{\! R} & 3&2&\frac 16\\
{u'_L} &  3 &1&\frac 23\\
{d'_L} & 3 &1&-\frac 13\\
\left( \begin{matrix} \nu'_e \\ e'\end{matrix} \right)_{\! R} & 1&2&-\frac 12\\
{e'_L} &  1 &1&-1
\end{matrix}\label{Rspectrum}
\ee
coupled to the gravitational metric and connection. In this multiplet the doublet $\left(\begin{matrix} u'\\ d'\end{matrix}\right)_R$  describes right-handed particles and left-handed antiparticles, while the singlets $u'_L$ and $d'_L$ represent left-handed particles and right-handed antiparticles, and similarly for the leptons. This multiplet couples to the $SU(3)_R\times SU(2)\times U(1)_R$ gauge fields. The corresponding kinetic part of the action is 
\be
S_f^{(+)}\equiv S_{fR}
&=&\int d^4 x \, \left(\sqrt{g}\, i\overline {{\psi'}}  
\gamma^a
 e_a^{\mu}
\left(\ED_\mu^{(+)}+\frac 12 \omega_\mu \right)\psi'\right)( x)\label{fermionaction+}
\ee
where $\psi'$ represents the right multiplet \eqref{Rspectrum}, and 
\be
\ED^{(+)}_\mu=\partial_\mu -i{\sfg}_X^+ X^{(+)}_\mu -i{\sfg}_WW_\mu -i{\sfg}_B^+B_\mu^{(+)} \label{XWQ+}
\ee
The symbols $X^{(+)}_\mu,W_\mu,B^{(+)}_\mu$ refer to the $SU(3)_{R}, SU(2)$ and $U(1)_{R}$ potentials. 

What is the reason for this doubling? As is well known the MSM defined on a trivial metric background is anomaly  free. However, when we couple it to a nontrivial metric it develops nontrivial trace anomalies with densities $ F\star F$ and $R\star R$, where $F$ is the curvature of the $SU(2)$ gauge field and $R$ is the gravitational curvature, \cite{LB,BLGS24}. The anomaly analysis of the mirror multiplet is the same as for the left one except for the sign of the anomalies which is reversed. In particular the trace anomaly due to the gauge field   $F\equiv F^{\mathsf su(2)}$ and to $R$ are the same, but with opposite sign. Therefore the overall sum of the obstructive anomalies of the system vanishes.
It is important to remarks that both multiplets couple to the same $SU(2)$  gauge fields. Only in this case do all obstructive anomalies cancel! 

It is opportune to insist on the main features of this theory:  both multiplets interact with the same metric and the same $SU(2)$ gauge fields, but with different $SU(3)\times U(1)$ gauge fields on the left and the right. This is what makes the overall theory free of obstructive anomalies. In other words the two sectors share the same weak and gravitational forces, while they have different and independent strong and hypercharge interactions. The full theory will be denoted simply by ${\cal T}={\cal T}_L \cup {\cal T}_R$. The symbol $\cup$ is to recall that the two half theories have the metric and the SU(2) gauge potentials in common.

In \cite{BG24,LB25} the pure gauge action and the pure gravity action, as well as the Higgs-like scalar sector and Yukawa couplings were explicitly written down. We dispense with it here. Of course we should consider three families of left-handed and three families of right-handed fermions, but for simplicity we shall understand the relative formulas here.

Let us conclude this presentation of the $\cal T$ theory with a comment on its physical interpretation.
It splits into two halves, each with distinct scalar and fermion matter components. The gauge groups $ SU(3)_L$ and $SU(3)_R$ as well as $ U(1)_L$ and $ U(1)_R$, respectively, are also distinct. Instead the metric and the $SU(2)$ gauge fields are the same on both sides. The left sector and the right sector interact with each other only via the latter fields and in no other way. For instance the right quarks (right-handed particles and left-handed antiparticles) $\left(\begin{matrix} u_R\\ d_R\end{matrix}\right)$   interact among themselves strongly via the $ SU(3)_R$ gauge bosons and electromagnetically via the $ U(1)_R$ potential. Moreover due to the Yukawa couplings they interact with the right doublet of scalars and also weakly via the $SU(2)$ gauge fields and gravitationally via the metric. However with the mediation of the latter they interact also with the right fermions (right-handed particles and left-handed antiparticles) $\left(\begin{matrix} u_L\\ d_L\end{matrix}\right)$ .

If the $\cal T$  theory is to represent the universe and its evolution we should complete it with one or more scalar fields to describe inflation and possibly dark energy. This can be easily done but again we omit it here for brevity and refer for instance to \cite{LB25,LB26}. Now the main question is: if the left part of $\cal T$ describes the MSM in interaction with gravity, what is the role of the right sector?  
The right part of the total matter+energy evolves in a way similar to the ordinary matter+energy, although with different coupling constants, masses and cosmological constant. At the present energy scale the interaction between the two reduces to the gravitational force, because the weak force between left and right matter, which is common to both, has a range which is too short to be effective at macroscopic distances. It is likely that the evolution in the right part  might be sensibly different from the ordinary matter concerning inflation, particle creation, density perturbations and so on. In summary, the idea is that there is another world (the right one) near to ours, which we cannot `see', but at most `feel' via gravity, and also via the weak force if the energy is high enough. It is of course very natural to interpret it as the dark matter world. And this is the way we understand the right sector in the continuation of our argument below.

\section{Massless neutrinos in a left-right symmetric SM}

The Godhaber experiment's neutrino  is a left-handed object with definite helicity (negative), denoted henceforth $\nu_{eL}$. It is natural to represent it by means of a left-handed Weyl spinor, which is necessarily massless. However  there is a problem: a Weyl spinor has definite energy and momentum, 0 mass and definite helicity, but it does not have a causal propagator. How can it propagate in spacetime? This is a serious problem not only from our physical representation, but also from a formal point of view, because without a propagator we cannot (perturbatively) quantize the theory containing Weyl fermions.

A way out can perhaps be found from the remark made in section 3, around eq.\eqref{Diracprop}, according to which a propagator exists if the Weyl spinor is accompanied by another Weyl spinor of opposite chirality, so as to constitute a Dirac fermion and this is the only possible causal propagator for fermions. In other words a Weyl fermion propagates only forming a couple with an opposite chirality partner.  This means a Weyl fermion is a full physical entity. But faced with the Goldhaber experiment this concept raises a puzzle: where is the right-handed partner of $\nu_{eL}$? The right handed partner $\nu_{eR}$ has never been found. Does the ${\cal T}$ theory of the previous section contain any useful information about it and explain why if it exist it is not visible to us? In the present section we would like to 
explore this possibility. 

In order to approach these question it is useful to recall how in the SM the Weyl fermions couple in such a way as to give rise to Dirac propagators.  Let us focus for instance on the quarks fields $u_L$ and $u_R$. The kinetic action \eqref{fermionaction-} comes from the doublet ${\sf d}_L= \left( \begin{matrix} u_L\\d_L\end{matrix}\right)$
\be
S_f^{(-)}\equiv S_{fL}
&=&\int d^4 x \,  \left(i\,\overline {{{\sf d}_L}}  
\gamma^\mu
\ED^{(-)}_\mu{\sf d}_L + i\,\overline {{u_R}}  
\gamma^\mu\ED^{(-)}_\mu\,u_R + \ldots\right)\0
\ee
where the dots refer to the remaining fields, and form now on we consider a flat metric. It is clear that this part of the action contains the kinetic action of a Dirac field $U= u_L+u_R$. Moreover the Yukawa couplings with the Higgs field can give rise to the mass term $\sim \overline {u_L} \, u_R + h.c.$. Massive or not, it is clear that the $U$ quark has a well-defined propagator of the Dirac type.
The same is true for the couples $d_L, d_R$ and $e_L, e_R$. But it is impossible to repeat this construction for $\nu_{eL}$, which is a genuine Weyl spinor, within the MSM. However looking at the spectrum of $\cal T$ we see that it is possible for the couple $\nu_{eL}, \nu'_{eR}$. Let us discuss this possibility and its consequences.

The kinematic part of the action for the couple $\nu_{eL}, {\nu}_{eR}'$ 
\be
S_f\equiv S_{fL}+S_{fR}
=\int d^4 x \,  \left(i\,\overline {{\nu_{eL}}}  
\gamma^\mu
\ED^{(\pm)}_\mu{\nu_{eL}} + i\,\overline {{\nu'_{eR}}}  
\gamma^{\mu}
\ED^{(\pm)}_\mu\nu'_{eR} + \ldots\right)\label{fermionactionuLuR}
\ee
In this and the previous equation
\be
\ED^{(\pm)}_\mu=\partial_\mu-i{\sfg}_WW_\mu -i{\sfg}_B^\pm B_\mu^{(\pm)}  \label{WQB}
\ee
This defines the kinetic part of a Dirac fermion $\nu^e=\nu_{eL}+\nu_{eR}$. As for the mass term see the comment below. As anticipated above we assume that $\nu^e$ is massless.

To start with let us make explicit the interaction terms involving $\nu_{eL}, e_L$, $ e_R,\nu'_{eR},e_L', e_R'$ (disregarding from now on the metric, since  in the sequel it does not play any role) They are
\be 
&&\,{\sf g}_W \left[ \frac 12 W^3_\mu \left(\overline {\nu_{eL}}\gamma^\mu \nu_{eL} -\overline {e_L }\gamma^\mu e_L \right)+ \frac {W^+_\mu }{\sqrt{2}} \overline {\nu_{eL}}\gamma^\mu e_L+  \frac {W^-_\mu }{\sqrt{2} }\overline  {e_L} \gamma^\mu \nu_{eL}\right]\0\\
&&+ \,{\sf g}_W \left[ \frac 12 W^3_\mu \left(\overline {\nu'_{eR}}\gamma^\mu \nu'_{eR} -\overline {e'_R }\gamma^\mu e'_R \right)+ \frac {W^+_\mu }{\sqrt{2}} \overline {\nu'_{eR}}\gamma^\mu e'_R+  \frac {W^-_\mu }{\sqrt{2} }\overline  {e'_R} \gamma^\mu \nu'_{eR}\right]\0\\
&&-\frac 1 2{\sf  g}^-_{B} B^{(-)}_\mu \left[\overline {\nu_{eL}}\gamma^\mu \nu_{eL}+\overline {e_L }\gamma^\mu e_L + \overline {e_R} \gamma^\mu e_R\right]\0\\
&&- \frac  1 2{\sf  g}^+_{B} B^{(+)}_\mu \left[\overline {\nu'_{eR}}\gamma^\mu \nu'_{eR}+\overline {e'_R }\gamma^\mu e'_R + \overline {e'_L} \gamma^\mu e'_L\right]\label{intterms}
\ee
There are in addition the standard interaction terms with a scalar Higgs doublet $H_-$ on the left and $H_+$ on the right side, which we dispense with here.

$\nu_{eL}$ has interaction terms with  $e_L, W_\mu^\pm$ and $B^{(-)}_\mu$. In turn $\nu'_{eR}$
directly interacts with $e_R', W_\mu^\pm$ and $B^{(+)}_\mu$.

Let us return now to the Golhaber experiment. The lowest order contribution to the process of electron capture with neutrino emission, \eqref{pennu}, is given by the interaction of an up quark in the nucleus which gets transformed to a down quark with emission of a W boson. The latter in turn interacts with an electron to emit a neutrino. In terms of Weyl spinor components this writes
\be
u_L + e_L \longrightarrow d_L + \nu_{eL}\label{uLeL}
\ee
which involves the following terms among those of \eqref{intterms}
\be
\,\frac {{\sf g}_W }{\sqrt 2}\left[ {W^+_\mu }\overline {\nu_{eL}}\gamma^\mu e_L+{W^-_\mu }\overline  {e_L} \gamma^\mu \nu_{eL}\right]+
\frac {{\sf g}_W }{\sqrt 2}\left[  {W^+_\mu }\overline {u_L}\gamma^\mu d_L+{W^-_\mu }\overline  {d_L} \gamma^\mu u_{L}\right]\label{int1}
\ee
from \eqref{fermionactionuLuR}, where the proton is schematically represented as the bound state $p=u_Ld_R u_L$ and the neutron as $n=u_Ld_Rd_L$.

From our discussion above we know that $\nu_{eL}$  exists as a classical solution of the Weyl-Dirac equation with definite energy, momentum and helicity, but it does not propagate. To define a propagator we need a right-handed companion $\nu'_{eR}$. This companion does not exist in the MSM, but it exists in this enlarged theory. For instance it might be produced in a process mirror to the previous one, which in terms of spinor fields read
 \be
u'_R + e'_R \longrightarrow d'_R + \nu'_{eR}\label{u'Re'R}
\ee
with the dark proton given by $u'_Rd'_Lu'_R$ and the dark neutro as $u'_Ld'_Ld'_R$. The lowest order contributions involve the terms
\be
\,\frac {{\sf g}_W }{\sqrt 2}\left[ {W^+_\mu }\overline {\nu'_{eR}}\gamma^\mu e'_R+{W^-_\mu }\overline  {e'_R} \gamma^\mu \nu'_{eR}\right]+
\frac {{\sf g}_W }{\sqrt 2}\left[  {W^+_\mu }\overline {u'_R}\gamma^\mu d'_R+{W^-_\mu }\overline  {d'_R} \gamma^\mu u'_{R}\right]\label{int2}
\ee

But we can have also the processes
\be
u_L + e_L \longrightarrow d'_R + \nu'_{eR}\label{uLeLd'R}
\ee
and
\be
u'_R + e'_R \longrightarrow d_L + \nu_{eL} \label{u'Re'R}
\ee

Whatever the process, let us suppose that the emissions of $\nu_{eL}$ and $\nu'_{eR}$ are simultaneous. Then they give rise to a propagating neutrino. The important remark is that $\nu'_{eR}$ cannot be detected (at least not with the same kind of experiment as  $\nu_{eL}$).
This is because in the case of \eqref{uLeLd'R} the process would give rise to an exotic neutron $u_Ld_L d'_R $. In the case of \eqref{u'Re'R} because the capture of $e'_R$ occurs entirely in the mirror sector. It should be kept in mind that, although the two sectors have the weak force in common (gravity does not play a role here), both reactions involve mirror quarks which have different strong and electromagnetic interactions than the SM counterparts. 

It should be added  that similar consideration hold for other weak processes like the free neutron decay $n\longrightarrow p + e + \overline {\nu_e}$. 
While waiting for technologies and experiments that may allow us to detect matter in the mirror sector, we can say that  $\nu'_{eR}$ `cannot be seen'.

There is an aspect in the above discussion that needs further elaboration: the formation of the `bound state' $\nu_e= \nu_{eL}+ \nu'_{eR}$. As was explained above only $\nu_e$ is able to propagate in spacetime. Therefore it $ \nu_{eL}$ needs a right-handed partner. The two partners are produced in separate events, but only when two Europium atoms are ready for the emission, one of $ \nu_{eL}$, the other of $\nu'_{eR}$, can the event of the electron capture with a neutrino emission take place. In the first case we see the recoil of $_{62} Sm^{*152}$, in the second case we do not see anything with the apparatus based on the SM particles and interactions. We must conclude that this coincidence presupposes a sort of action at a distance, a simultaneous stimulation of two atoms not due to a force but to something similar to what happens in the entanglement phenomena: an action-at-a-distance principle that underpins the physical significance of the concept of Weyl fermions. This principle must be at work also when $\nu_{eL}$ is absorbed in reverse reactions.

A point of the previous argument is still in the shade. A Weyl spinor has been defined before a incomplete physical object and now we claim that what we see in the Goldhaber et al. experiment has all the features of a Weyl spinor: is there a sense in which we can speak of a Weyl spinor as  a definite physical object? To explain this issue we must recall the process of measurement in  quantum mechanics: measuring an observable maps the system into an eigenstate of the observable; the eigenstate is not the entire system but only a part of it. In the case of  our interest the detection of a neutrino in the Goldhaber et al. experiment corresponds to measuring chirality. There are two eigenstates of it corresponding to two Weyl spinors with opposite chirality. At this stage we can finally give a clear physical sense to a Weyl fermion: it is not a full fledged particle, but only the output of the measurement of chirality.

\section{Conclusion}

The message we wish to convey in this paper is that Goldhaber et al. experiment poses serious problems to the theoretical physics of weak interactions. That experiment detected a left-handed neutrino with negative helicity. This implies it is massless. The standard mathematical interpretation is that it can be represented by a Weyl spinor, excluding right away the possibility of a Majorana spinor because the latter has indefinite chirality. However, as it has been shown above, based on \cite{LBRS26}, it is not possible to define a causal propagator for a Weyl spinor field, with the consequence that a Weyl spinor field is an incomplete physical object. It becomes a fully defined physically acceptable object when coupled to another Weyl spinor of opposite chirality, to form a Dirac spinor, which is well known to propagate.  But, supposing that the Goldhaver experiment's  neutrino can be represented by a Dirac spinor, how comes that we are able to detect only half of it? In this paper we have tried to clarify this puzzle by embedding the physical system of the experiment in an enlarged SM, which consists of two copies of the SM, one with the familiar chirality called left and another with reverse chirality, the right.The two halves share the same weak and gravitational  force, but have distinct strong and electromagnetic interactions. We have shown that in this theory there is room for a plausible interpretation of the fateful experiment, which, incidentally, allow us to clarify what a Weyl spinor represents. Let us add that the overall level of our proposal is qualitative.  Even accepting the hypothesis that the mirror sector does represent dark matter, the couplings and in general the physics of the mirror part of the theory are not known on the basis of the present experimental knowledge. But we think the qualitative picture we have shown above is attractive and the hypothesis underneath it worth delving into. 

\section*{Acknowledgements}
 I woud like to thank the organizers of the workshop {\it What comes beyond the standard model} 2026, for inviting me to  give a talk which is at the origin of this comment. 
\vskip 0,6cm

\end{document}